\documentclass{appolb}
\usepackage{epsfig}
\newcommand{\beq}{\begin{equation}}
\newcommand{\eeq}{\end{equation}}
\newcommand{\un}{\underline}
\newcommand{\as}{\alpha_s}
\def\eq#1{{(\ref{#1})}}
\def\fig#1{{Fig.~\ref{#1}}}

\begin{document}
\title{Manifestation of the Color Glass Condensate\\
in particle production at RHIC
\thanks{Presented at XXXIV International Symposium on Multiparticle 
Dynamics. July 26 - August 1, 2004. Sonoma County, California, USA }
}
\author{Kirill Tuchin
\address{Physics Department, Brookhaven National Laboratory,\\
Upton, NY 11973-5000, USA}
}
\maketitle
\begin{abstract}
We discuss general properties of the Color Glass Condensate. We show that    
predictions for particle production in p(d)A and AA collisions  derived 
from these properties are in agreement with data collected at RHIC.  
\end{abstract}
\PACS{13.85.Hd,13.60.Hb,13.85.Ni,24.85}
  
In this paper we discuss the experimental signatures of the new form 
of nuclear matter -- the Color Glass Condensate (CGC) in particle 
production at RHIC. Let us first see how the notion of the CGC arises in 
pA collisions at high energy. Consider a process of inclusive particle 
production in pA collisions in a nucleus rest frame. At high energies 
the typical values of the Bjorken $x$ are small. It is well-known 
that at small $x$ hard processes develop over large `coherence length' 
$l_c$. In particular, a gluon production is coherent over $l_c\simeq 1/(M 
x)$, where $M$ is the proton's mass. For instance, at midrapidity at RHIC 
the coherence 
length of 2~GeV gluon is $l_c\simeq 20$~fm (at $\sqrt{s}=200$ GeV). It is 
much 
bigger than the size of the target $\simeq 6.5$ fm. This allows formal 
separation of the gluon production process into two parts: slow gluon 
emission described by the proton's light cone wave function, and almost 
instantaneous interaction with the target at given impact 
parameter $\un b$ described by the amplitude $N_G(\un r, \un b, x)$, where 
$\un r$ is the variable Fourier-conjugated to the gluon's transverse 
momentum $\un k$. In the one gluon exchange approximation, assuming that 
scattering on different nucleons 
is independent,  one arrives at the formula \cite{MV,kjklw,KM}
\beq\label{ngscat}
N_G(\un r,\un b,x)=1-\exp\left[-\frac{1}{4}\un r^2 Q_s^2 S(\un b) 
\ln(1/r\mu)\right],
\eeq
similar to the  Glauber formula for the low energy hadron-nucleus 
scattering. Here $Q_s$ is a parameter with dimension of mass, $S(\un 
b)$ is a nuclear profile function  
and $r\equiv |\un r|$. $N_G(\un r,\un b,x)$ can be interpreted as a 
forward scattering amplitude of a gluon dipole off a heavy nucleus. One 
can see 
from \eq{ngscat} that for hard gluons, such that  $1/r\sim k \gg Q_s$ 
the scattering amplitude coincides with the usual perturbative expression 
\beq\label{ngpert}
N_G^\mathrm{pert}(\un r,\un b,x)= \un r^2\, \pi^2\,\as\,
\rho\,T(\un b)\,xG(r,x)/ (2 \,C_F).
\eeq
In the opposite limit $k\ll Q_s$ 
the scattering amplitude \eq{ngscat} becomes independent of its variables 
$\un r$, $\un b$ and $x$ as it approaches its unitarity limit. This is the 
phenomenon of \emph{saturation} in nuclear 
and hadronic reactions \cite{GLR}. The scale $Q_s$ is called  \emph{the 
saturation scale}. 
Eq.~\eq{ngpert} implies that $Q_s^2\propto A^{1/3}$, where $A$ is the 
atomic number. Thus, for a very big nucleus $A\gg 1$ the saturation scale 
becomes a perturbative scale $Q_s\gg \Lambda_\mathrm{QCD}$ which in turn 
means that $\as(Q_s^2)\ll 1$.  It can be 
argued that the total multiplicity of produced gluons is dominated by 
gluons with the typical momentum $k\simeq Q_s$. Therefore, gluon 
production in pA collisions at high energies is calculable in the 
perturbation theory.
 
Note, that the scattering amplitude \eq{ngscat} was calculated in a 
quasi-classical approximation. The quasi-classical approximation 
corresponds to a quantum system with high occupation numbers. In terms of 
the QCD action $S_\mathrm{QCD}$ this implies $S_\mathrm{QCD}\gg 1$. At 
small $x$ gluons dominate over quarks. Therefore, we have
\beq\label{act}
\frac{1}{g^2}\int
d^4x\, \mathrm{tr}\, \tilde G_{\mu\nu}(x)\, \tilde G^{\mu\nu}(x)\gg 1,
\eeq
where the rescaled gluon field is defined as $\tilde A_\mu^a(x)=g\, 
A_\mu^a(x)$. On the other hand, $\as(Q_s^2)\ll 1$. Therefore, the typical 
gluon field of nucleus is of order of $A_\mu^a\sim 1/g$ \cite{MV}. This 
strong 
gluon field at small coupling is called \emph{the Color Glass Condensate}. 
This configuration is very much different from the perturbation theory where 
both the gluon (and quark) field and the coupling are small, and from the  
non-perturbative regime where both gluon (and quark) field and the 
coupling are large \cite{KL}. Phenomenologically, the CGC in a 
quasi-classical approximation manifests itself  as a saturation 
of the scattering amplitude $N_G(\un r,\un b,x)$ at small transverse 
momenta \cite{GLR}. 

The CGC in a quasi-classical approximation can be thought of as a model of 
 multiple rescatterings of a hadron in a heavy nucleus at high energy. As such it 
has much in common with many 
other models of multiple rescatterings. In particular, their common 
prediction is the Cronin effect, i.\ e.\ enhancement of particle 
production at intermediate transverse momenta $k$ in pA collisions as 
compared with pp scaled by the atomic number A. The origin of this effect 
is simple: a gluon traversing a heavy nucleus gains additional transverse 
momentum due to multiple rescatterings. On the other hand, in a 
quasi-classical approximation the total number of particles is conserved. 
Therefore, if there are less particles with low transverse momentum, 
then there are more particles with high transverse momentum 
\cite{KKT,Alb,BKW,IIT}. Of course, this effect is predicted to 
increase 
for heavier nuclei or more central 
collisions. 

However, a quasi-classical approximation breaks down at high energy since 
quantum evolution becomes an important process. Indeed, additional gluon 
production is parametrically of order $\as\ln(1/x)$. Therefore, 
at $x \ll e^{1/\as}$ a quasi-classical approximation is no longer valid.
 One might 
attempt to take the evolution into account using collinear factorization, 
which basically means incoherent  production of gluons in the proton's 
wave function. As 
a result, proton will suffer more scatterings in a nucleus and 
the Cronin effect will increase with energy/rapidity. However,   
this expectation contradicts the experimental data as we discuss later, 
see \fig{fig2}. The reason is that the collinear factorization scheme 
and the OPE break down as soon as multiple rescatterings are important, see 
e.\ g.\ \cite{LMT}. 
This is because each additional scattering is a higher-twist effect. 
The effect of coherence of the parton evolution at high energies can be  
taken into account in the \emph{nonlinear evolution equations} of QCD 
\cite{GLR,BK,JKLW}. These 
equations describe the  high energy quantum evolution of the CGC. That is, 
if the scattering amplitude $N_G(\un r,\un b, x)$ is known at some initial 
value of $x_0$, e.\ g.\ as given by \eq{ngscat}, the evolution equations 
allow calculation of the scattering amplitude at any $x<x_0$. 
 
In the large $N_c$ approximation the differential cross section for a 
gluon production can be written in the $k_T$-factorized form 
\cite{KM,KT,Braun}
\beq\label{paev}
\frac{d \sigma^{pA}_G}{d^2 k \ dy} \, = \, \frac{C_F \, S_A \, S_d}{\as \,
\pi \, (2 \pi)^3} \,
\frac{1}{{\un k}^2} \, \int d^2 r \,
\nabla^2_z \, n_G ({\un z}, Y-y)
 \, e^{- i {\un k}
\cdot {\un r}} \, \nabla^2_r \, N_G ({\un r}, y),
\eeq
where $y=\ln(1/x)$ and $n_G(\un r, \un b, y)$ is a forward gluon dipole 
scattering  amplitude off a proton. $S_A$ and $S_p$ are cross 
sectional areas of the gold nucleus and proton correspondingly and 
$Y$ is 
the total rapidity interval.
The evolution effects in a nucleus are 
enhanced by  a factor of $A^{1/3}\gg 1$ as compared to those in proton 
(deuteron). 
Therefore, $n_G(\un r, \un b, y)$ approximately satisfies the linear 
BFKL evolution equation \cite{BFKL} (this is correct at not very high 
energies, 
when the Pomeron loops are small). The gluon dipole scattering amplitude 
can be related to the unintegrated gluon distribution function $\phi(\un 
k,x)$ as \cite{KT}
\beq\label{distr}
\phi(\un k, x)=\frac{C_F}{\as (2\pi)^3}\int d^2b\, d^2r\, e^{-i\un 
k\cdot 
\un r}\,\nabla_r^2 \, N_G(\un r,\un b, x). 
\eeq
The main property of $\phi(\un k,x)$ which follows directly from the 
nonlinear evolution equations is the  \emph{geometric scaling} which 
means that  $\phi(\un k,x)=\phi(k/Q_s(x))$, i.\ e.\ the gluon distribution 
becomes a function of only one variable at low $x$ \cite{GLR}! Here 
$Q_s(x)$ is 
the same saturation scale as in \eq{ngscat}. Being the only dimensional 
parameter at low $x$, $Q_s(x)$ sets the scale for the gluon field.  
Eq.~\eq{act} implies $A_\mu^a\sim Q_s(x)/g$. In course of  evolution 
$A_\mu^a$ increases as $x$ decreases due to increase of number of color 
sources. Hence, $Q_s(x)$ is increasing function of $1/x$. It follows 
from the nonlinear evolution equation that \cite{LT}:
\beq\label{satscale}
Q_s^2(x)=\left(\frac{x_0}{x}\right)^\lambda\, A^{1/3}\,\mathrm{GeV}^2.
\eeq
where $\lambda\approx 0.3$ \cite{Trian}. The same gluon distribution 
function $\phi(\un 
k,x)$ \eq{distr} enters expressions for the structure functions  in Deep 
Inelastic 
Scattering. It allows to fit the initial value of $x$ using experimental 
data collected at HERA. 
In \eq{satscale}: $x_0=3\cdot 10^{-4}$ and $\lambda= 0.28$. For 
RHIC and LHC it is convenient 
to write \eq{satscale} in the center-of-mass frame 
\beq
Q_s^2=\left(\frac{\sqrt{s}}{3.3\, \mathrm{TeV}}\right)^\lambda\,e^{\pm 
\lambda y}\, 
A^{1/3}\, \mathrm{GeV}^2.
\eeq

\begin{figure}
\begin{tabular}{cc}
\epsfig{file=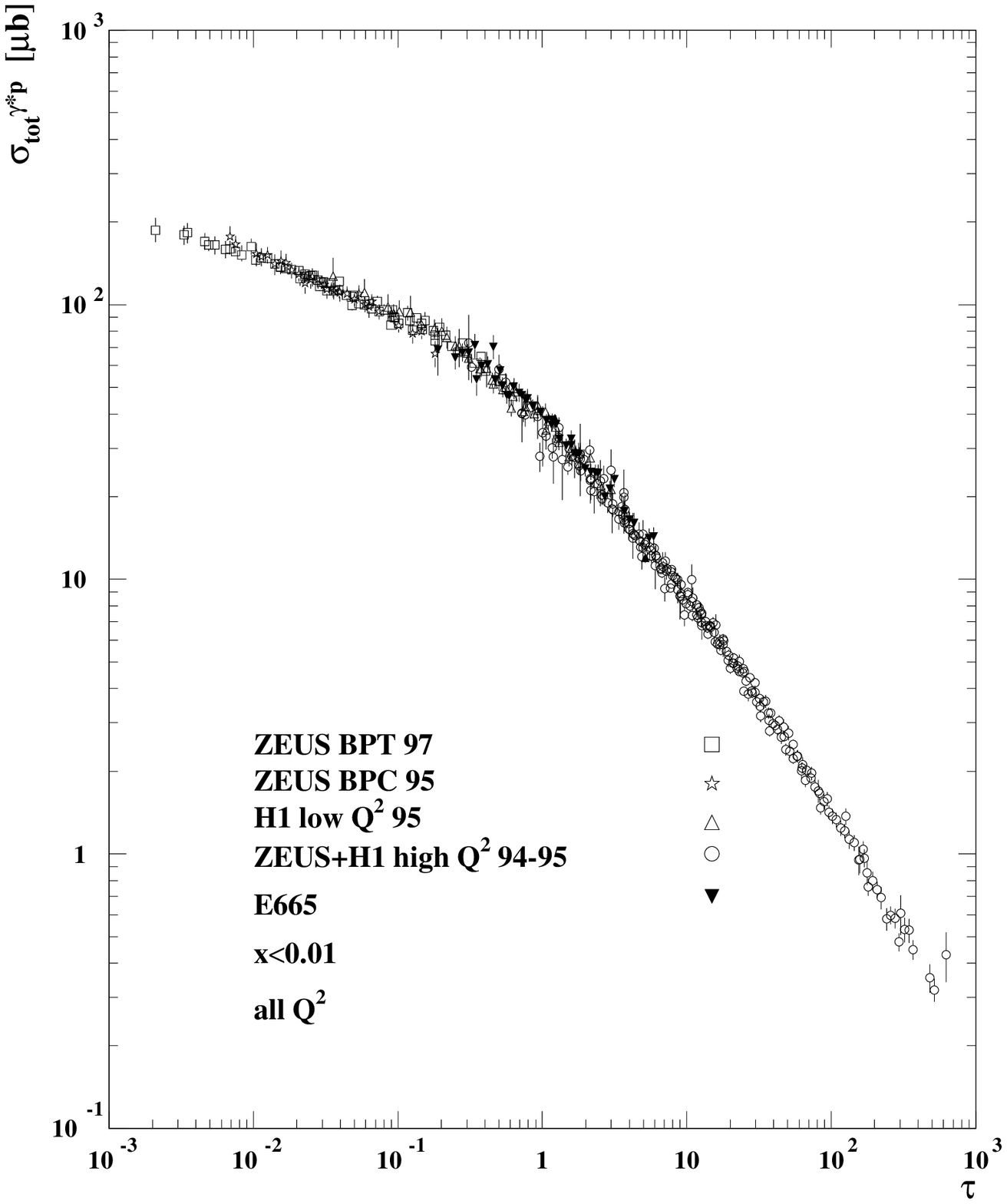,width=7cm}&
\epsfig{file=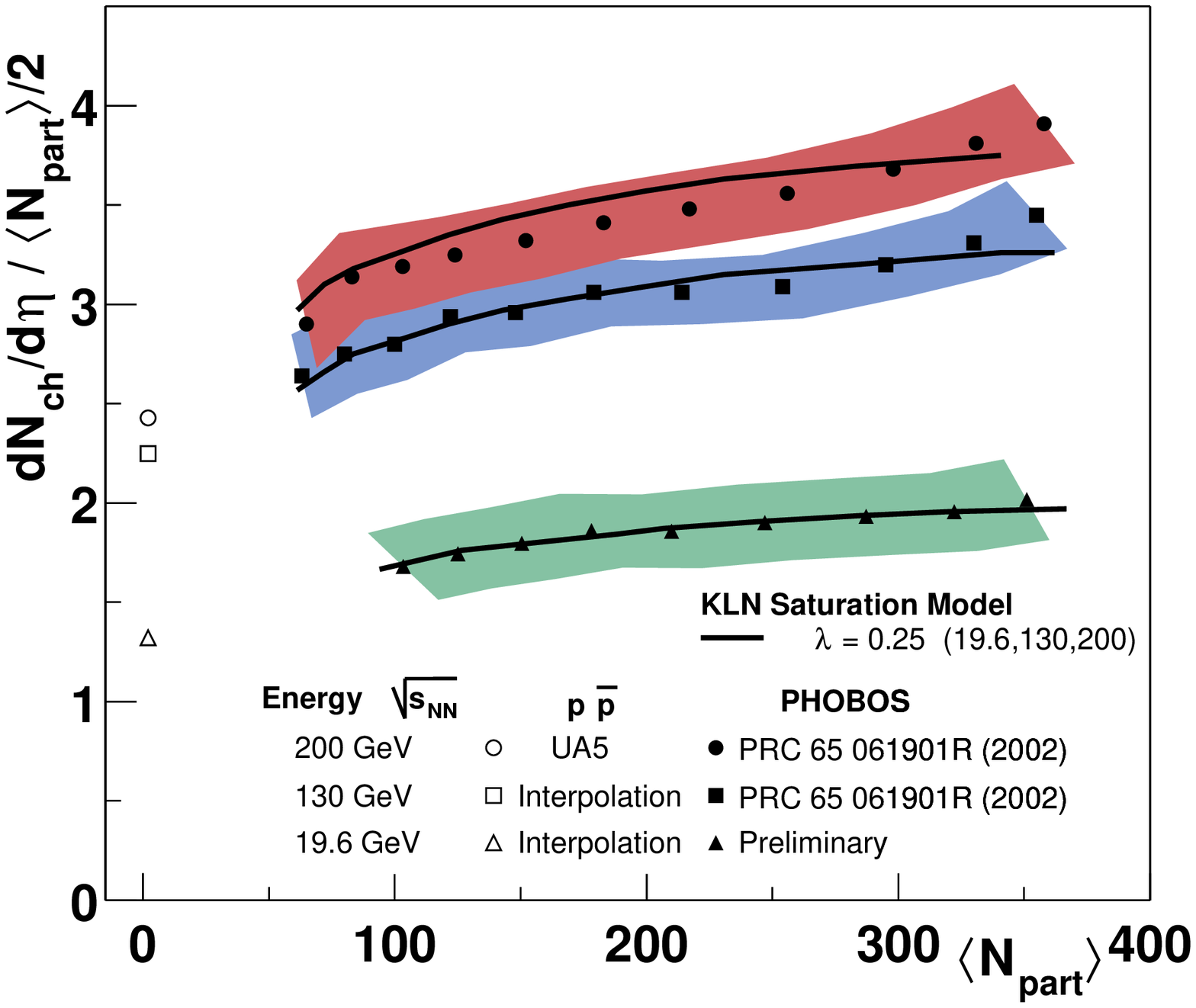,width=5.5cm,height=7cm}\\
(a) & (b)
\end{tabular}
\caption{(a) Geometric scaling of the total DIS cross section 
$\sigma(Q^2,x)$ as a function of $\tau=Q^2/Q_s^2$\cite{geom}. (b) 
Geometric scaling  of the total charged hadron multiplicity in Au-Au 
collisions at RHIC. 
Solid lines: prediction of the saturation model of \cite{KL}. Prediction  
based on the collinear factorization is quite different $dN/d\eta\, 
N_\mathrm{part}^{-1}\propto N_\mathrm{part}^{1/3}$. }\label{fig1}
\end{figure}

\begin{figure}
\begin{tabular}{cc}
\epsfig{file=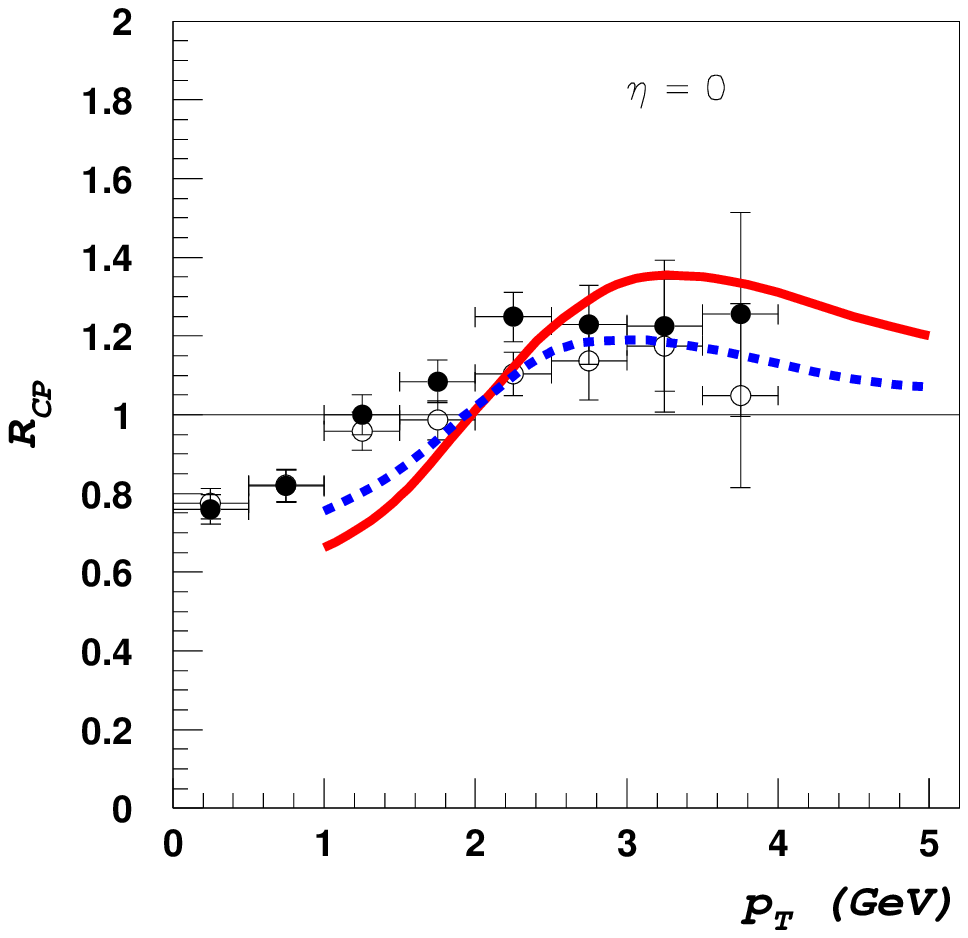,width=5.5cm,height=4.5cm}&
\epsfig{file=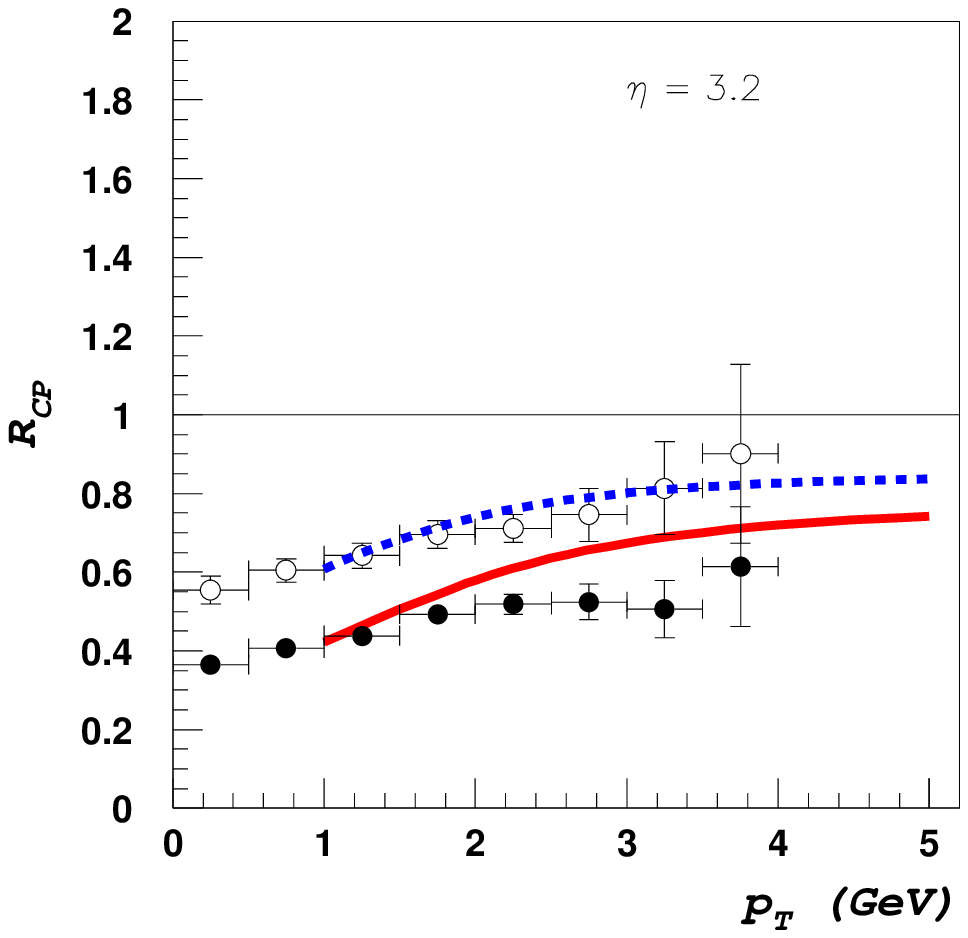,width=5.5cm,height=4.5cm}
\end{tabular}
\caption{The nuclear modification factor: ratio of the charged 
hadron multiplicity  in central (full dotes, $b\approx 3$ fm) and 
semi-central (open dots, $b\approx 5$ fm) dA collisions to those in peripheral 
ones($b\approx 7$ 
fm) rescaled by the ratio of corresponding numbers of participated 
nucleons. Lines: result of a simple CGC model of \cite{KKT}.  
}\label{fig2}
\end{figure}

The geometric scaling of the gluon distribution holds as long as the 
logarithms of energy gained in course of the BFKL evolution are bigger 
than the logarithms of transverse momentum gained in course of the DGLAP 
evolution:
\beq
\as \log Q_s^2/\,\Lambda^2 \,\sim \,\as\, y\,\gg \, \as\, \log \un 
k^2/Q_s^2,
\eeq 
which implies the geometric scaling in a wide kinematical region 
$k<k_\mathrm{geom}=Q_s^2/\Lambda$ \cite{IIM}.
The experimental evidences of the geometric scaling in DIS and heavy ion 
collisions are shown in \fig{fig1}.

Another consequence of the quantum evolution on the unintegrated gluon 
distribution \eq{distr} is that its anomalous dimension $\gamma$ acquires 
strong 
dependence on the scaling variable $k/Q_s$. In the perturbative regime 
$k\gg k_\mathrm{geom}$ we get the usual leading-twist expression $\phi(\un 
q,x)\propto S_A \,Q_s^2/\un q^2$  modulo DGLAP corrections. However, in 
the saturation $k<Q_s$, $\phi(\un q,x)\propto S_A$, i.\ e.\  
$\gamma\rightarrow 
0$. As we have already noted this signals the breakdown of the OPE. In the 
intermediate region $Q_s<k<k_\mathrm{geom}$ the saddle point of the 
BFKL amplitude is located at $\gamma\approx 1/2$, which implies $\phi(\un
q,x)\propto S_A Q_s/q$. Recalling, that $Q_s\propto A^{1/3}$ we 
find  that at $k< k_\mathrm{geom}$ the gluon distribution in a nucleus of 
atomic 
number $A$ is less than $A$ times the gluon distribution in  a proton: 
\beq\label{suprho}
\frac{\phi_A(k,x)}{A\,\phi_p(k,x)}=\frac{1}{A^\rho},
\eeq
with $\rho=1/3$ at $k\ll Q_s$ and $\rho\approx 1/6$ at 
$Q_s<k<k_\mathrm{geom}$. Let me emphasize that  
the CGC takes into account two nuclear shadowing effects. First one is a 
quasi-classical effect of multiple rescatterings. It necessarily requires 
higher twist effects to be included in a calculation. It predicts 
suppression at $k<Q_s$ followed by enhancement at $k\sim Q_s$. Second one 
is a quantum evolution effect. It predicts suppression in wide kinematical 
region $k<k_\mathrm{geom}$ both in the linear evolution region at $k>Q_s$ 
(`leading twist shadowing') and the nonlinear evolution one 
at $k<Q_s$ (saturation).  
Using \eq{paev} and \eq{distr} it is easily seen that nuclear shadowing in 
$\phi_A(k,x)$ translates into suppression of particle production in  
deuteron-gold collisions \cite{KLM,KKT,Alb,BKW,IIT}.
In \fig{fig2} recent RHIC data  for 
charged particle production at different rapidities and centralities 
\cite{BRAHMSdata} is 
shown. We see  that at $\eta=0$ there is the Cronin enhancement of the 
particle production in dA as compared to pp in central collisions as 
predicted by the CGC as well as by many multiple rescatterings models 
\cite{AG}. 
This implies that quasi-classical approximation is valid. At 
pseudo-rapidity $\eta=3.2$ corresponding to 25 times smaller $x$'s  
than at $\eta=0$ for the same 
transverse momentum, the evolution becomes essential. It manifests itself 
as suppression of  particle production at large transverse momenta in 
p(d)A as compared to pp at higher rapidities and centralities. 

Not only that none of the existing conventional nuclear shadowing models 
can explain \emph{both} the Cronin effect and the suppression in 
the deuteron fragmentation region, but also none of those shadowing models can 
explain 
the large value of that suppression \cite{GSV}. In the framework of CGC 
both effects 
are \emph{predicted} to follow from  the nonlinear evolution 
equation \cite{KLM,KKT,Alb,BKW}. The value 
of suppression factor comes naturally as a consequence of \eq{suprho}. 

Since the suppression of charged particles in p(d)A and AA at 
forward rapidities at RHIC originates in a gluon shadowing \eq{suprho} 
there should be similar suppression in open charm production. In that case 
the region of the geometric scaling, and hence of suppression,  is 
$m_t<k_\mathrm{geom}$, where $m_t^2=\un k^2+m_c^2$ \cite{DKKT}. 
Another important signature of CGC is  weakening of jet--jet 
correlations \cite{correl}. Indeed, since at low $x$ a lot of particles 
with 
a typical momentum $Q_s$ can be produced 
in single nucleon--nucleon subcollision any two of them need not to 
be correlated back-to-back unless their transverse momenta are very 
large. 
Once all these pieces of evidence all collected together they will become 
a strong evidence for the Color Glass Condensate at RHIC. 

This research was supported by the U.S. Department of Energy under
Contract No. DE-AC02-98CH10886.
%%%%%%%%%%%%%%%%%%%%%%%%%%%%%%%%%%%%%%%

\end{document}